\documentclass{moriond}

\usepackage{amsmath}
\usepackage{amsthm}

\usepackage{units}
\usepackage{mathrsfs}
\usepackage{amssymb}
\usepackage{slashed}
\usepackage{multirow}
\usepackage{mathtools}
\usepackage{nicefrac}

\newcommand{\A}{\mathcal{A}}
\newcommand{\U}{\mathbf{U}}
\newcommand{\T}{\mathbf{T}}
\newcommand{\V}{\mathbf{V}}
\newcommand{\LL}{\mathscr{L}}
\newcommand{\F}{\mathcal{F}}
\newcommand{\bb}{b} %b parameter in the SShh coupling
\newcommand{\s}{\sigma}

\DeclareMathOperator{\Tr}{Tr}
\newcommand{\tr}{\Tr}
\renewcommand{\to}{\rightarrow}
\newcommand{\de}{\partial}

\begin{document}
\vspace*{4cm}
\title{NON-LINEAR HIGGS PORTAL TO DARK MATTER}

\author{ ROC\'IO DEL REY BAJO }

\address{ Departamento de F\'isica Te\'orica UAM and Instituto de F\'isica Te\'orica UAM-CSIC, \\ Universidad Aut\'onoma de Madrid, Cantoblanco, 28049, Spain}

\maketitle\abstracts{
The Higgs portal to scalar Dark Matter is considered in the context of non-linearly realised electroweak symmetry breaking. We determine the interactions of gauge bosons and the physical Higgs particle $h$ to a scalar singlet Dark Matter candidate $S$ in an effective description. The main phenomenological differences \textit{w.r.t.} the standard scenario can be seen in the Dark Matter relic abundance, in direct/indirect searches and in signals at colliders.
}
%%%%%%%%%%%%%%%%%%%%%%%%%%%%%%%%%%%%%%%%%%%%%%%%
\section{Motivation}
%%%%%%%%%%%%%%%%%%%%%%%%%%%%%%%%%%%%%%%%%%%%%%%%
Dark Matter (DM) cannot be explained within the Standard Model (SM); its presence is one of the experimental evidences for physics beyond the SM. The nature of the Higgs particle  also raises a quandary, as the electroweak (EW) hierarchy problem remains unsolved. The lightness of the Higgs  may result from its being a pseudo-Goldstone boson (GB) of a global symmetry~\cite{Kaplan:1983fs}, spontaneously broken by strong dynamics at a high scale $\Lambda_s\gg v$, as typically arises in scenarios where electroweak symmetry breaking (EWSB) is non-linearly realised. Much as the interactions of QCD pions are weighted down by the pion decay constant and described by an effective field theory (EFT) with a derivative ordering,  those of the EW pseudo-Goldstone bosons -- the longitudinal components of the $W^\pm$ and $Z$ plus the $h$-- will be weighed down by $f$ ($\Lambda_s\leq 4\pi f $)\cite{Manohar:1983md}. An EFT approach is adopted to avoid the specificities of particular models.

%%%%%%%%%%%%%%%%%%%%%%%%%%%%%%%%%%%%%%%%%%%%%%%%
\section{Standard \textit{vs.} non-linear Higgs portals}
%%%%%%%%%%%%%%%%%%%%%%%%%%%%%%%%%%%%%%%%%%%%%%%%

The Higgs portal~\cite{Patt:2006fw} is one of the three possible renormalisable ($d\le4$) interactions between the SM and DM (along with vector-like and fermion portals). Assuming a $Z_2$ symmetry~\cite{Silveira:1985rk,Veltman:1989vw}, under which  $S$ is odd and the SM fields are even for DM stability, the \textit{standard} portal is defined by
\begin{equation}
\lambda_{S} S^2\Phi^\dag \Phi \longrightarrow \lambda_{S} S^2 (v + h)^2 \longrightarrow \lambda_{S} S^2 (2vh + h^2)\,,
\label{SMHportal}
\end{equation} 
shown in unitary gauge, with $\Phi$ the $SU(2)_L$ Higgs doublet, $h$ the physical Higgs particle, $v$ the EW scale as defined from the Fermi decay constant and $\lambda_{S}$ the Higgs portal coupling.

%%%%%%%% RELIC DENSITY PLOTS %%%%%%%%%%%%%%%%%
\begin{figure}co
\centering
\begin{minipage}{0.4\linewidth}
\centerline{\includegraphics[width=1.2\linewidth]{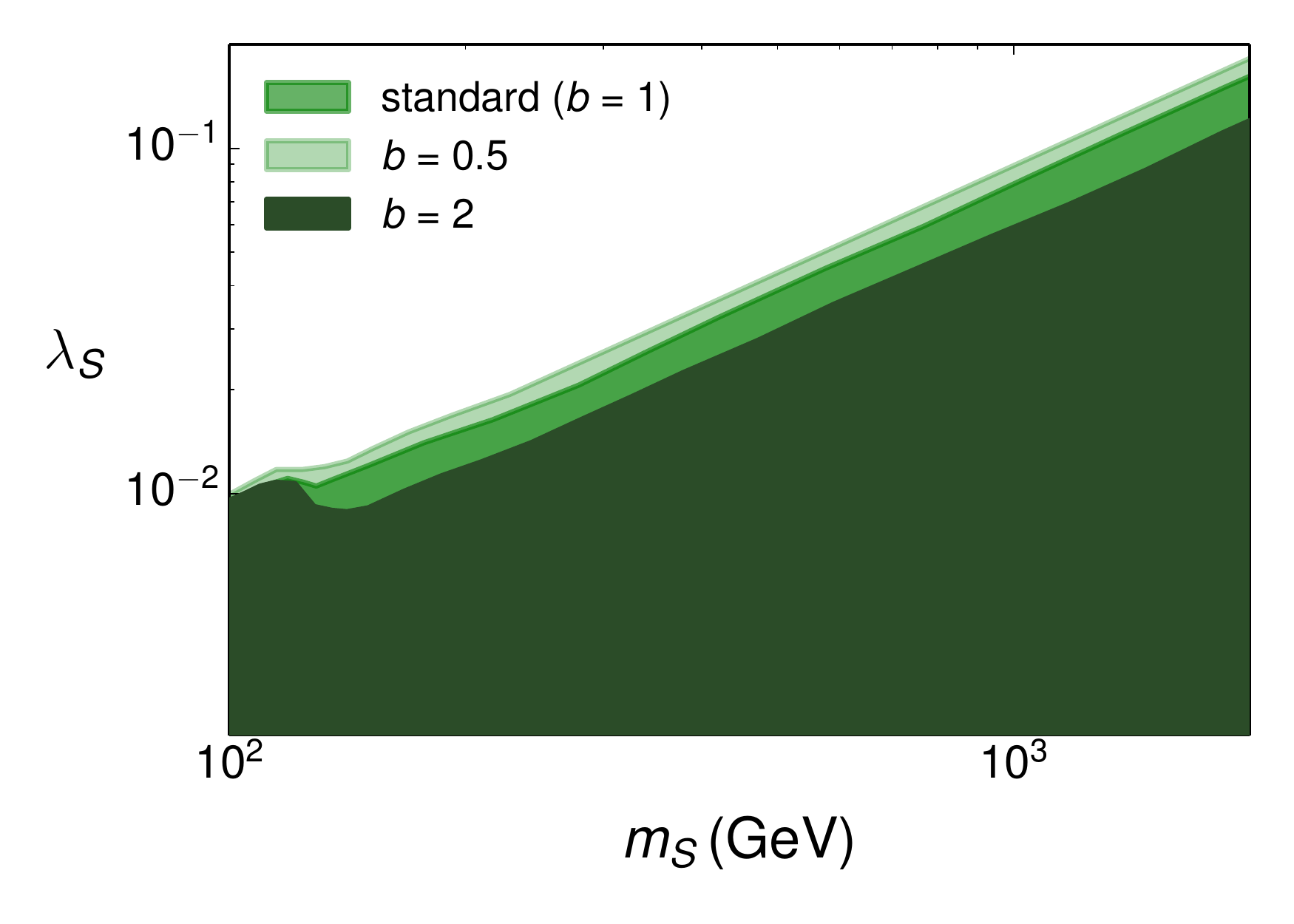}}
\end{minipage}
\hspace{2cm}
\begin{minipage}{0.4\linewidth}
\centerline{\includegraphics[width=1.2\linewidth]{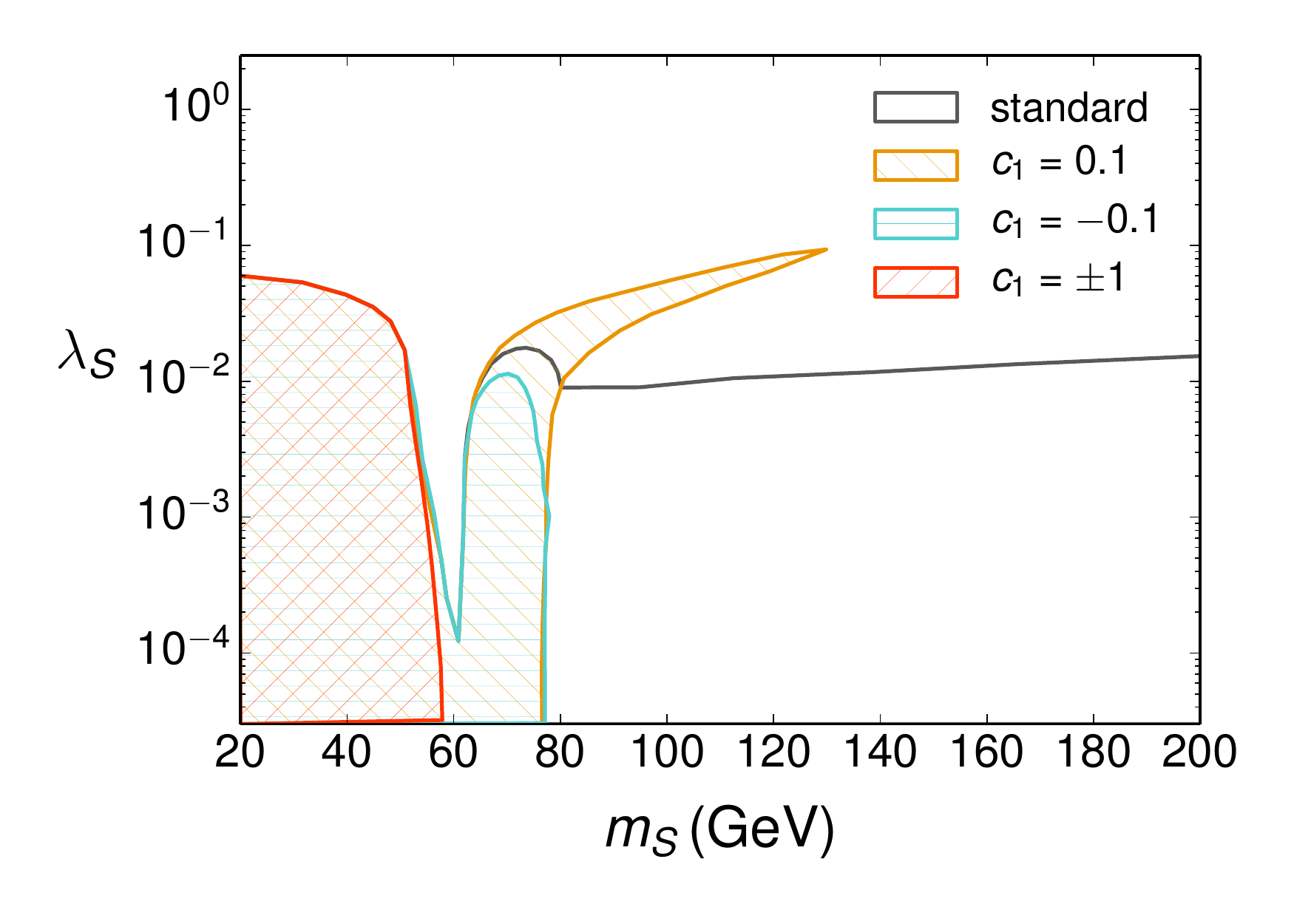}}
\end{minipage}
\caption[]{Regions in the $(m_S, \lambda_S)$ plane excluded by the condition $\Omega_S h^2 \leq 0.12$ for different values of $\bb$ (left) and in the presence of non-linear operator $\A_1$ for different values of $c_1 \in [-1,1]$ (right).}
\label{fig:relic_plots}
\end{figure}
%%%%%%%%%%%%%%%%%%%%%%%%%%%%%%%%%%%%%%%

In non-linear scenarios\cite{Contino:2010rs,Alonso:2012px}, the physical Higgs field may no longer behave as an exact EW doublet at low energies. It can be treated effectively as a generic SM scalar singlet with arbitrary couplings.  The typical SM dependence on $(v+h)$ is to be replaced by a generic polynomial
\begin{equation}
\F(h)= 1 + 2a\frac{h}{v}+ b \left(\frac{h}{v}\right)^2 + \dots\,.
\end{equation}
In addition, the interactions of the physical $h$ particle are not necessarily correlated with those of the  $W^\pm$ and $Z$ longitudinal components, denoted by  $\pi(x)$ in the unitary GB  matrix:
\begin{equation}
\U(x)\equiv e^{i\sigma_a \pi^a(x)/v}\,. \label{Udef}
\end{equation}
Note that the ``pions" here are suppressed by $v$, where the natural GB weight is in fact the scale $f$; this encodes the fine-tuning affecting these models.
While in linear BSM scenarios $h$ and $\U(x)$ are parts of the same object $\Phi$, they are treated independently in the non-linear setup.

In the effective non-linear Lagrangian, only the leading terms weighted down by $\Lambda_{DM}$ and $\Lambda_{s}$ (both $\Lambda_s,\, \Lambda_{DM} \gg f \gg v$) are kept, which  
means no explicit dependence on them. It can  be written as  $\LL=\LL_{EW}+\LL_S$. Several choices are possible for the EW leading order Lagrangian $\LL_{EW}$, although this is of minor impact here (see Ref.~\cite{Brivio:2015kia}).
$\LL_S$  encodes the DM interactions~\cite{Brivio:2015kia}:
\begin{equation}
\LL_S=\frac{1}{2}\de_\mu S \de^\mu S- \dfrac{m_S^2}{2}  S^2 
-\lambda_{S}S^2\left(2vh+\bb h^2\right)+\sum_{i=1}^5 c_i\A_i(h) +\dots 
\label{L0S}
\end{equation}
where the $\A_i$ operators form a basis \footnote{$\A_3-\A_5$  contain sources of custodial symmetry breaking further than those present in the SM (hypercharge in this case). The contribution of $\A_4$ to the $Z$ mass  vanishes while that from $\A_5$ arises only at the two loop level, and no significant constraint on their operator coefficient follows the $\rho$ parameter and  EW precision data. These observables do receive a one-loop contribution from $\A_3$, implying a bound estimated to be around $c_3 \sim 0.1$. }:
\begin{equation}
\begin{matrix}
\begin{aligned}
\A_1&=\tr(\V_\mu\V^\mu)S^2\F_1(h)\\
\A_2&=S^2 \square \F_{2}(h)\\
\A_3&=\tr(\T\V_\mu)\tr(\T\V^\mu ) S^2 \F_{3}(h)\\  \end{aligned}\hspace{1cm}&\begin{aligned}
\A_4&=i\tr(\T\V_\mu) (\de^\mu S^2) \F_{4}(h)\\
\A_5&=i\tr(\T\V_\mu)S^2\de^\mu \F_{5}(h)\,.
\end{aligned}
\end{matrix}
\label{scalar.op}
\end{equation} 
The dots in Eq.~(\ref{L0S}) stand for terms with more than two $h$ bosons and/or more than two $S$ fields, which are not phenomenologically relevant in the analysis below and are henceforth discarded. 

%%%%%%%% SUMMARY PLOTS %%%%%%%%%%%%%%%%%
\begin{figure}
\centering
\begin{minipage}{0.4\linewidth}
\centerline{\includegraphics[width=1.15\linewidth]{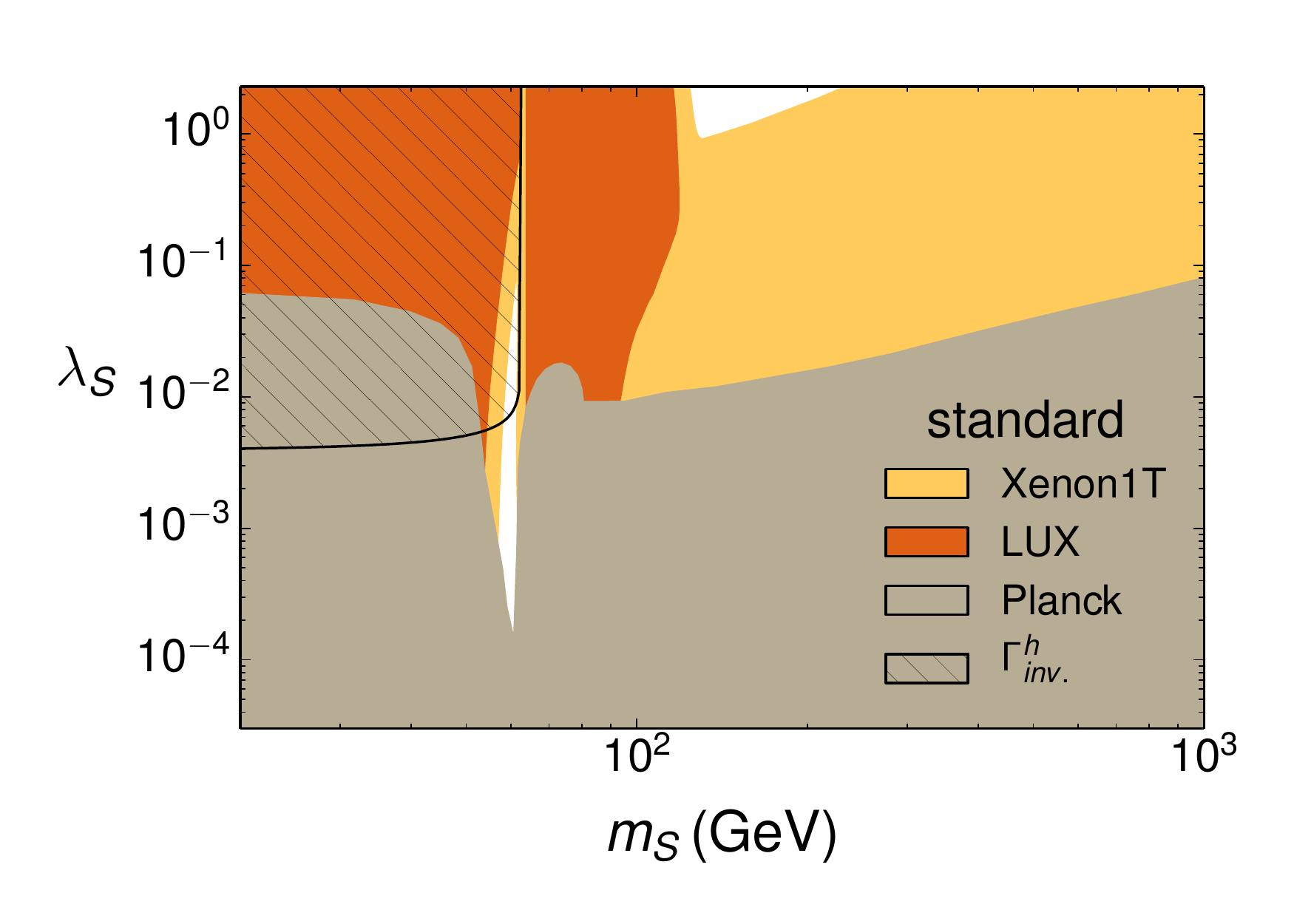}}
\end{minipage}
\hspace{2cm}
\begin{minipage}{0.4\linewidth}
\centerline{\includegraphics[width=1.15\linewidth]{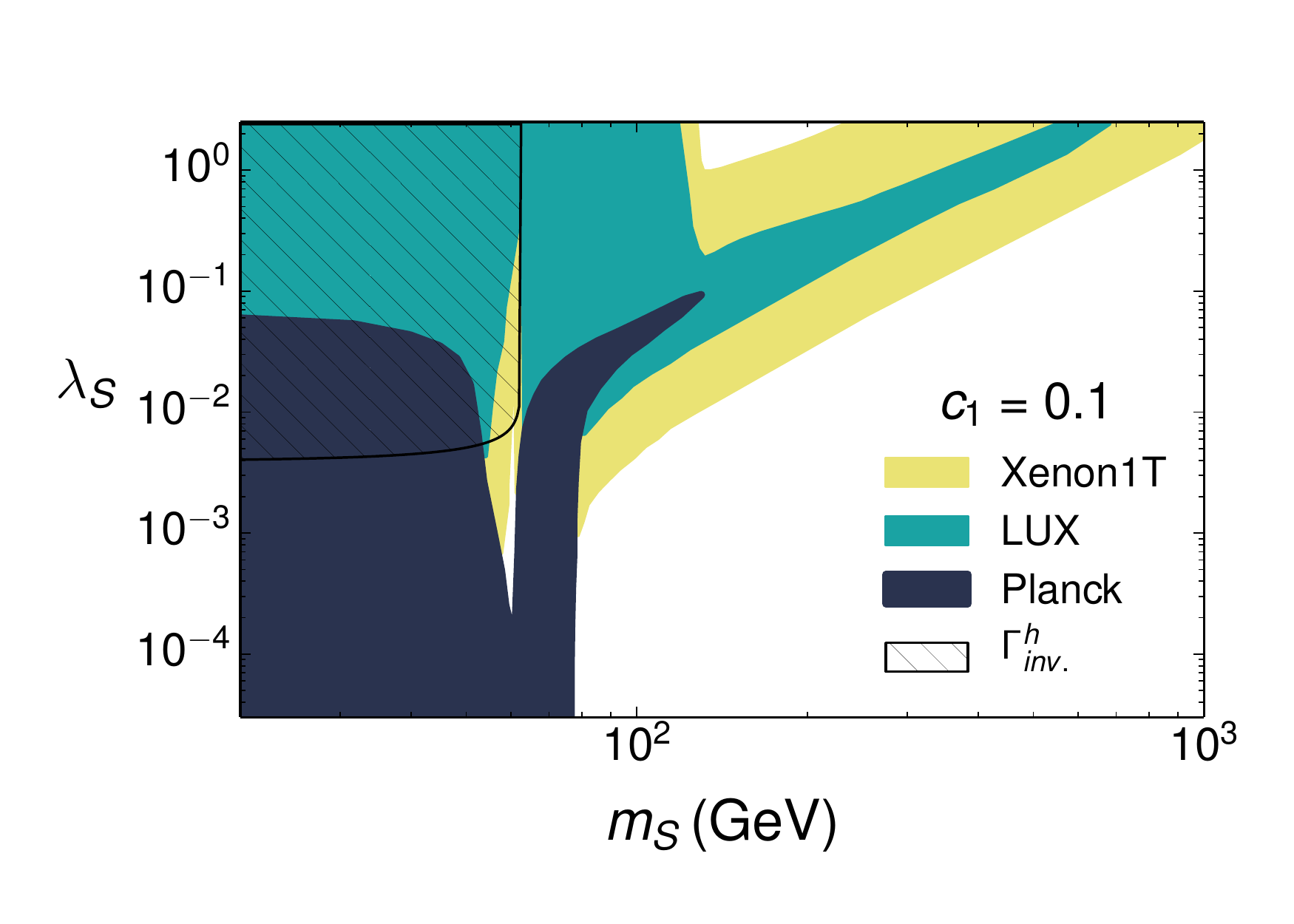}}
\end{minipage}
\caption[]{Standard portal (Left) and non-linear portal in the presence of operator $\A_1$ with $c_1=0.1$ (Right) in the $(m_S,\lambda_S)$ plane. Regions excluded by current bounds from Planck (brown/ dark blue), LUX (orange/teal) and invisible Higgs decay (hatched) are plotted together with the projected reach of XENON1T (yellow/lime). }
\label{fig:summary_plots}
\end{figure}
%%%%%%%%%%%%%%%%%%%%%%%%%%%%%%%%%%%%%

%%%%%%%%%%%%%%%%%%%%%%%%%%%%%%%%%%%%%%%%%%%%%%%%%%%%
\section{Dark Matter phenomenology}
%%%%%%%%%%%%%%%%%%%%%%%%%%%%%%%%%%%%%%%%%%%%%%%%%%%%
We showcase some salient features of non-linear Higgs portal scenarios varying one coefficient of \{$b,\, c_i$\} at a time, and confront them with the standard portal ($\bb=1,\, c_i=0$). This allows to single out the impact of each effective operator ensuring a clear and conservative  comparison.

\textbf{Dark Matter relic density:}\quad Assuming  $S$ to be a thermal relic, its abundance $\Omega_S$ is determined by the thermally averaged annihilation cross section into SM particles times relative velocity in the early Universe $(\sigma \mathrm{v})_\text{ann}=\sigma(SS\rightarrow XX)\,\mathrm{v}$ .
We require the abundance not to exceed the observed value~\cite{Ade:2015xua}, assuming $S$ may either be the sole DM particle or a member of a larger DM sector: $\Omega_{S} h^2 \leq \Omega_{\mathrm{DM}} h^2 \simeq 0.12$. Deviations from the SM-like correlation between $SSh$ and $SShh$ couplings (described by values $\bb \neq 1$) have important consequence for $m_S>m_h$ through the process $SS\rightarrow hh$. Values of $\bb>1$ enhance this annihilation, shrinking the excluded region (see Figure~\ref{fig:relic_plots} (left)). Non-linear operators  $\A_i$ affect DM annihilations into gauge bosons, Higgses and $b$-quarks. 
Interactions induced by $\A_1$  modify the annihilation  into two gauge bosons (relevant for $m_S \gtrsim 65$ GeV). For $c_1<0$, interference with the linear term increases $\s_\text{ann}$, making some previously excluded points viable  (Figure~\ref{fig:relic_plots} (right)). If $c_1>0$ the interference is destructive: new cancellations  exclude previously allowed points (\textit{e.g.} the yellow ``branch" for $c_1=0.1$).

\textbf{Dark Matter direct detection (DD):}\quad DM-nucleon interactions occur in our scenario via Higgs exchange and, in the non-linear case via $W^\pm$ and $Z$ exchange. The strongest bounds constrain the spin-independent cross section $\sigma_{\mathrm{SI}}$ for $S$ on nucleons. Again, $S$ may be a member of a larger DM sector, in which case DD bounds are to be rescaled: $\sigma_{\mathrm{SI}}(S\,N\to S\,N)\times (\Omega_S/\Omega_{\mathrm{DM}})\leq 
\sigma_\text{exp}^{lim}$.
The current~\cite{Akerib:2013tjd}, and projected~\cite{Aprile:2012zx} DD exclusion regions are shown in Figure~\ref{fig:summary_plots} for the standard Higgs portal scenario, and  in the presence of the non-linear operator $\A_1$ with  $c_1=0.1$. While  $\A_1$ doesn't affect the $S$-nucleon scattering to first approximation ($SSZZ$ and $SSW^{+}W^{-}$ vertices  do not enter the scattering at tree level), its impact on $\Omega_S$ affects the DD exclusion regions due to the necessary rescaling. The allowed parameter space is  enlarged for $c_1<0$ while for $c_1>0$  the exclusion region may  stretch further into an area that is allowed in the standard setup.
%%%%%%%%%%%%%%%        MONOZ/W PLOTS      %%%%%%
\begin{figure}\centering
\begin{minipage}{0.33\linewidth}
\centerline{\includegraphics[width=1.15\linewidth]{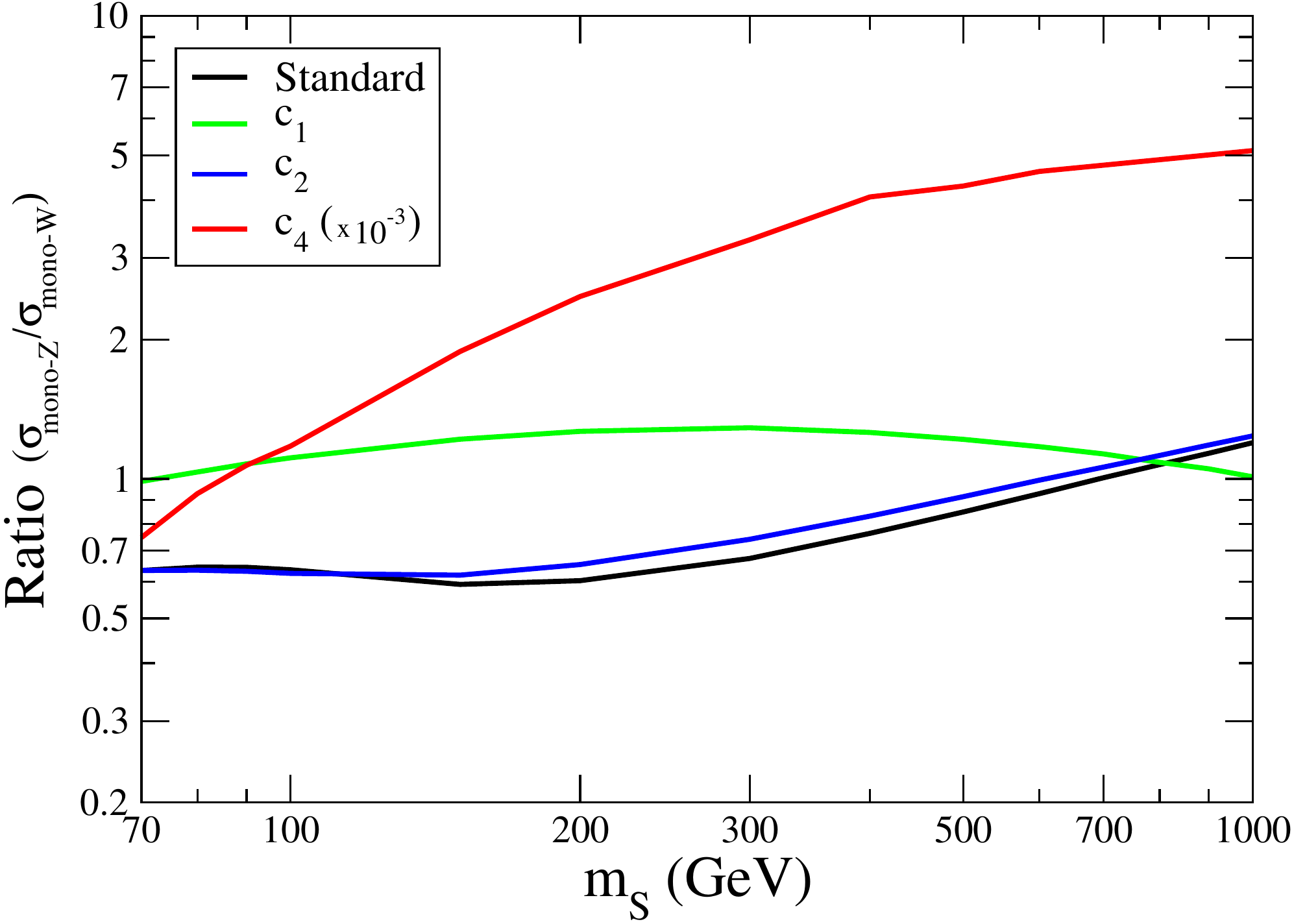}}
\end{minipage}
\hspace{2.5cm}
\begin{minipage}{0.32\linewidth}
\centerline{\includegraphics[width=1.15\linewidth]{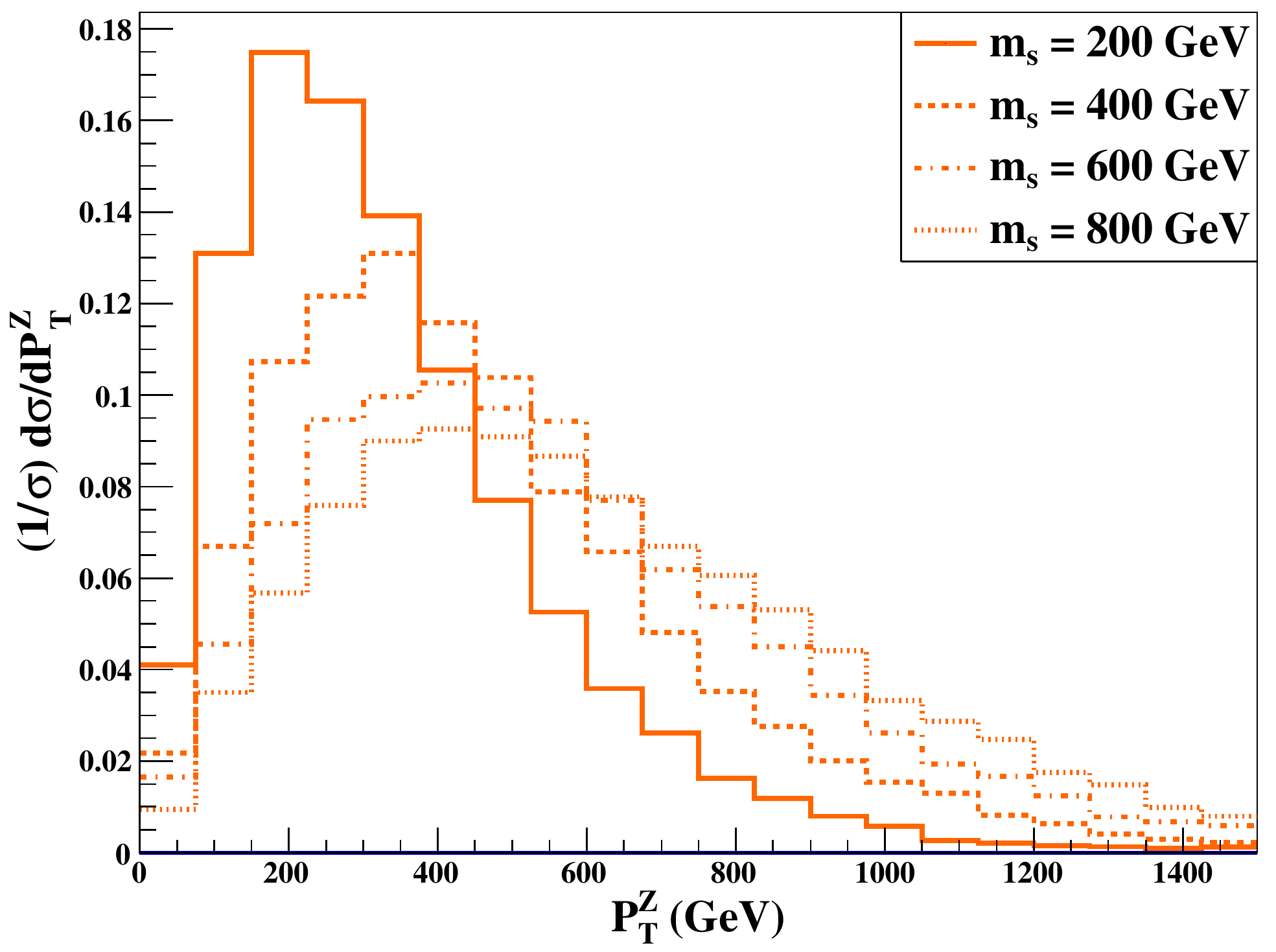}}
\end{minipage}
\caption[]{Left: Cross section ratio $R_{WZ}\equiv \sigma(pp\rightarrow ZSS)/\sigma(pp\rightarrow W^\pm SS)$ at $\sqrt{s}=13\,\text{TeV}$ as a function of $m_S$ in the standard Higgs portal (black line) and for different non-linear operators (colour). Right: Normalised differential $P_T^Z$ distributions for $pp\rightarrow ZSS$ for $\A_5$ for different DM masses. }
\label{fig:monoZW_plots}
\end{figure}
%%%%%%%%%%%%%%%%%%%%%%%%%%%%%%%%%%%%

\textbf{Invisibles Higgs decay width:}\quad The decay channel $h\to SS$ is open for $m_S<m_h/2$, contributing to the  Higgs invisible width $\Gamma_\text{inv}$. The presence of $\A_2$ has a significant impact: even for $\lambda_S \to 0$, $\Gamma_\text{inv}\neq0$ for $c_2 a_2 \neq 0$. We require $\text{BR}_\text{inv} = \Gamma_\text{inv}/(\Gamma_\text{inv}+\Gamma_\text{SM} )< 0.23$~\cite{Aad:2015pla}. While the presence of $\A_1$ does not modify this constraint (see Figure~\ref{fig:summary_plots}), that of $\A_2$ is not illustrated but should be commented: even for 
small values of this coefficient practically all the region $m_S<m_h/2$ is excluded.

\textbf{Dark matter at the LCH:}\quad A key probe of DM  at colliders are ``mono-$X$'' signatures --associated production of DM particles with a visible object $X$, which recoils against missing transverse energy $\slashed{E}_T$. 
The presence of non-linear Higgs portal interactions $\A_{1-5}$ has a dramatic impact, allowing EW production of DM via couplings to vector bosons, leading to mono-$W$, mono-$Z$ and mono-Higgs signatures with rates ${\cal O}(10^{1-4}) \times c_i^2$ bigger than the standard Higgs. A promising smoking gun consists of using  the ratio $R_{WZ} \equiv \s(pp\to Z SS)/\s(pp\to W^{\pm} SS)$, as shown in Figure~\ref{fig:monoZW_plots} (left). The impact of each non-linear operator determined in general independently of the value of the coefficient (either $c_i$ or $\lambda_S$ in the standard case). The ratio $R_{WZ}$ is a powerful discriminator for the cases of $\A_1$ and $\A_4$ (green and red curves respectively), and also trivially for $\A_5$, for which the mono-$W^{\pm}$ process is absent and $R_{WZ}\equiv\infty$. We show~\cite{Brivio:2015kia} how these effects could be alternatively explained by unnaturally large values of $d=6$ operator coefficients in a linear expansion. It is in principle possible to infer the DM mass from the mono-$X$   processes through the differential information on transverse momentum $P_T^{X}$  (\textit{e.g.} Figure~\ref{fig:monoZW_plots} (right)). The hypothetical observation of mono-$Z/W$ signals would allow to simultaneously extract a measurement of  $R_{WZ}$ and of $m_S$, identifying a unique point (surrounded by a finite error region) in the parameter space of Figure~\ref{fig:monoZW_plots} (left). Naively, the further away this point lies from the black line, the more disfavoured the standard portal scenario will be. 

%%%%%%%%%%%%%%%%%%%%%%%%%%%%%%%%%%%%%%%%%%%%%%%%
\section{Conclusions}
%%%%%%%%%%%%%%%%%%%%%%%%%%%%%%%%%%%%%%%%%%%%%%%%
In summary, a more general scenario of scalar Higgs portals, with non-linearly realised EWSB   gives rise to remarkable effects. Deviations from the SM-like correlations between i) single- and di-higgs couplings and ii) the interactions of $h$ and the longitudinal $W^\pm$ and $Z$ \textit{d.o.f.} deeply affect constraints on the parameter space in  the non-linear Higgs portal. Predictivity, however, is not lost as the appearance of new couplings and novel kinematic features at the renormalisable level provides handles to disentangle non-linear behaviour from the standard Higgs portal at colliders. In particular, we have  proposed observables that are able to distinguish the non-linear portal from the standard one (more details can be found in Ref.~\cite{Brivio:2015kia}). The search for Dark Matter and the quest for the nature of EWSB are major present challenges. We have discussed their interplay within an effective approach in the framework of the Higgs Dark Matter portal.

%%%%%%%%%%%%%%%%%%%%%%%%%%%%%%%%%%%%%%%%%%%%%%%%
\section*{Acknowledgements}
%%%%%%%%%%%%%%%%%%%%%%%%%%%%%%%%%%%%%%%%%%%%%%%%

My work is supported by CiCYT through the project FPA2012-31880, by the Spanish MINECO's ``Centro de Excelencia Severo Ochoa" Programme under the grant SEV-2012-0249 and and by  INVISIBLES-PLUS (H2020-MSCA-RISE-2015-690575). A special thanks to all the collaborators from Madrid (I. Brivio, M.B. Gavela and L.Merlo) and Sussex (K. Mimasu, J.M. No and V. Sanz). I thank the the organisers for the kind invitation and for such an enjoyable meeting.

%%%%%%%%%%%%%%%%%%%%%%%%%%%%%%%%%%%%%%%%%%%%%%%%%%%%
\section*{References}
%%%%%%%%%%%%%%%%%%%%%%%%%%%%%%%%%%%%%%%%%%%%%%%%%%%%
%\bibliography{biblio}{}

\begin{thebibliography}{10}
\footnotesize
\bibitem{Kaplan:1983fs}
David~B. Kaplan and Howard Georgi.
%\newblock {$SU(2)$ $\times$ $U(1)$ Breaking by Vacuum Misalignment}.
\newblock {\em Phys.Lett.}, B136:183, 1984.


\bibitem{Manohar:1983md}
Aneesh Manohar and Howard Georgi.
%\newblock {Chiral Quarks and the Nonrelativistic Quark Model}.
\newblock {\em Nucl.Phys.}, B234:189, 1984.

\bibitem{Appelquist:1980vg}
Thomas Appelquist and Claude~W. Bernard.
%\newblock {Strongly Interacting Higgs Bosons}.
\newblock {\em Phys. Rev.}, D22:200, 1980.

\bibitem{Longhitano:1980iz}
Anthony~C. Longhitano.
%\newblock {Heavy Higgs Bosons in the Weinberg-Salam Model}.
\newblock {\em Phys. Rev.}, D22:1166, 1980.

\bibitem{Longhitano:1980tm}
Anthony~C. Longhitano.
%\newblock {Low-Energy Impact of a Heavy Higgs Boson Sector}.
\newblock {\em Nucl.Phys.}, B188:118, 1981.

\bibitem{Patt:2006fw}
Brian Patt and Frank Wilczek.
%\newblock {Higgs-field portal into hidden sectors}.
\newblock  hep-ph/0605188, 2006.

\bibitem{Silveira:1985rk}
Vanda Silveira and A.~Zee.
%\newblock {Scalar Phantoms}.
\newblock {\em Phys. Lett.}, B161:136, 1985.

\bibitem{Veltman:1989vw}
M.~J.~G. Veltman and F.~J. Yndurain.
%\newblock {Radiative Corrections to W W Scattering}.
\newblock {\em Nucl. Phys.}, B325:1, 1989.
%
%\bibitem{Englert:1964et}
%F.~Englert and R.~Brout.
%%\newblock {Broken Symmetry and the Mass of Gauge Vector Mesons}.
%\newblock {\em Phys.Rev.Lett.}, 13:321--323, 1964.
%
%\bibitem{Higgs:1964ia}
%Peter~W. Higgs.
%%\newblock {Broken Symmetries, Massless Particles and Gauge Fields}.
%\newblock {\em Phys.Lett.}, 12:132--133, 1964.
%
%\bibitem{Higgs:1964pj}
%Peter~W. Higgs.
%%\newblock {Broken Symmetries and the Masses of Gauge Bosons}.
%\newblock {\em Phys.Rev.Lett.}, 13:508--509, 1964.

\bibitem{Contino:2010rs}
Roberto Contino.
%\newblock {The Higgs as a Composite Nambu-Goldstone Boson}.
\newblock arXiv:1005.4269, 2010.

\bibitem{Alonso:2012px}
R.~Alonso, M.B. Gavela, L.~Merlo, S.~Rigolin, and J.~Yepes.
%\newblock {The Effective Chiral Lagrangian for a Light Dynamical `Higgs'}.
\newblock {\em Phys.Lett.}, B722:330--335, 2013.



%\bibitem{Brivio:2016fzo}
%I.~Brivio, J.~Gonzalez-Fraile, M.~C. Gonzalez-Garcia, and L.~Merlo.
%%\newblock {The complete HEFT Lagrangian after the LHC Run I}.
%\newblock  arXiv:1604.06801, 2016


\bibitem{Brivio:2015kia}
I.~Brivio, M.~B. Gavela, L.~Merlo, K.~Mimasu, J.~M. No, R.~del Rey, and
  V.~Sanz.
%\newblock {Non-linear Higgs portal to Dark Matter}.
\newblock {\em JHEP}, 04:141, 2016.

\bibitem{Ade:2015xua}
P.~A.~R. Ade et~al.
%\newblock {Planck 2015 results. XIII. Cosmological parameters}.
\newblock 2015.

\bibitem{Akerib:2013tjd}
D.~S. Akerib et~al.
%\newblock {First results from the LUX dark matter experiment at the Sanford
%  Underground Research Facility}.
\newblock {\em Phys. Rev. Lett.}, 112:091303, 2014.

\bibitem{Aprile:2012zx}
Elena Aprile.
%\newblock {The XENON1T Dark Matter Search Experiment}.
\newblock {\em Springer Proc. Phys.}, 148:93--96, 2013.

\bibitem{Aad:2015pla}
Georges Aad et~al.
%\newblock {Constraints on new phenomena via Higgs boson couplings and invisible
%  decays with the ATLAS detector}.
\newblock {\em JHEP}, 11:206, 2015.

\end{thebibliography}
%\bibliographystyle{unsrt}
%\end{document}
%
%
%

\end{document}